# Efficient InGaN-based Red Light-Emitting Diodes by Modulating Trench Defects

*Zuojian Pan, Zhizhong Chen,\* Haodong Zhang, Han Yang, Chuhan Deng, Boyan Dong, Daqi Wang, Yuchen Li, Hai Lin, Weihua Chen, Fei Jiao, Xiangning Kang, Chuanyu Jia, Zhiwen Liang, Qi Wang,\* Guoyi Zhang, Bo Shen*

Z. Pan, Z. Chen, H. Zhang, H. Yang, C. Deng, B. Dong, D. Wang, Y. Li, H. Lin, W. Chen, F. Jiao, X. Kang, G. Zhang, B. Shen
State Key Laboratory for Artificial Microstructure and Mesoscopic Physics
School of Physics
Peking University
Beijing 100871, China
E-mail: zzchen@pku.edu.cn

Z. Chen, Z. Liang, Q. Wang, G. Zhang
Dongguan Institute of Optoelectronics
Peking University
Dongguan 523808, China
E-mail: wangq@pku-ioe.cn

Z. Chen, B. Shen
Yangtze Delta Institute of Optoelectronics
Peking University
Nantong 226000, China

F. Jiao
State Key Laboratory of Nuclear Physics and Technology
School of Physics
Peking University
Beijing 100871, China

C. Jia
School of International Academy of Microelectronics
Dongguan University of Technology
Dongguan 523000, China

**Table of Contents**

Trench defects have been innovatively utilized to achieve high-efficiency InGaN-based red light-emitting diodes (LEDs). The red quantum wells exhibit significant luminescence enhancement and wavelength redshift by appropriately modulating the trench defects. Red InGaN LEDs with an internal quantum efficiency of 16.4% are achieved by this method.

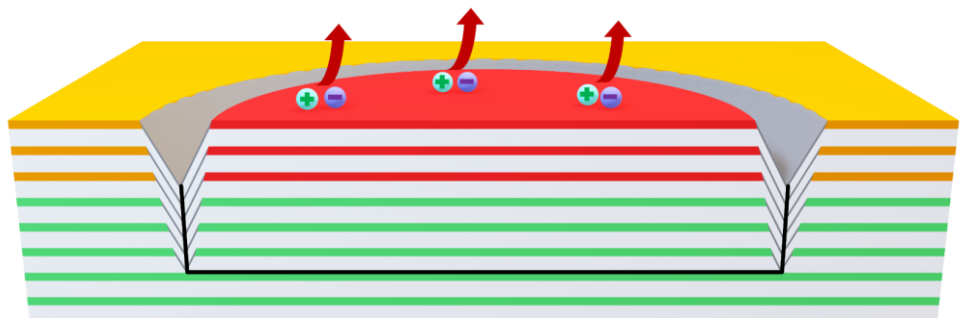

**Abstract**

Trench defects in multi-quantum wells (MQWs) have been considered as flawed structures that severely degraded the internal quantum efficiency (IQE) of light-emitting diodes (LEDs) in the past. In this research, trench defects are innovatively utilized to enhance the efficiency of red InGaN LEDs. Specifically, dual-color MQWs structures were applied to modulate trench defects. The upper red MQWs, grown on top of green MQWs with a high density of trench defects, exhibit a significant wavelength redshift of 68 nm and approximately 6-fold luminescence enhancement compared to those without intentionally introduced trench defects. Red InGaN LEDs with an IQE of 16.4% are achieved with this epitaxy growth strategy. Such wavelength redshift is attributed to the more indium incorporation due to the strain relaxation effect of trench defects. Moreover, the luminescence enhancement originates from the strong emission of the red MQWs inside trench defects, mainly attributed to strain relaxation and defect shielding by the wide and deep trenches. Achieving red emission by modulating trench defects is simple and reproducible without requiring additional substrate designs, which provides a novel way toward high-efficiency red InGaN LEDs.

# 1. Introduction

The full-color display of InGaN-based micro-light-emitting diodes (micro-LEDs) is crucial for emerging applications in virtual/augmented reality, ultrahigh-resolution mobile displays, and wearable displays[1-4]. While blue and green micro-LEDs achieve relativity high efficiencies, they remain very low in the red region[5]. The low efficiency of red InGaN LED is mainly attributed to the poor crystalline quality due to low-temperature growth[6-8] and a significant compressive strain introduced by lattice mismatch[9-11] in the active region. The low-temperature grown multi-quantum wells (MQWs) exhibit high-density defects, significantly reducing the radiative recombination efficiency[7, 8]. Moreover, severe compressive strain will induce strong polarizing electric fields in the InGaN quantum wells (QWs), leading to quantum-confined Stark effects (QCSE)[9]. Additionally, the compressive strain in the InGaN layer could resist the In incorporation due to the immense strain energy required for In-N bond, as suggested by the principle of free energy minimization[12, 13]. Previous research has pointed out various microstructures in MQWs that can suppress defect-related recombination and alleviate compressive stress[14], such as V-pits[15-17] and InGaN quantum dots[18, 19]. Notably, highly efficient red InGaN LEDs have been achieved by modulating V-pits[16, 17]. Thus, it is feasible to improve the efficiency of red InGaN LEDs by properly modulating the microstructures in MQWs.

Trench defects commonly occur inside InGaN/GaN MQWs grown by various methods[20-30]. The formation of trench defects is mainly related to the high In components and low growth temperatures of MQWs[23, 31, 32]. Previous studies show that trench defect originates from basal plane stacking faults (BSF) at the InGaN/GaN interface in MQWs[29, 33]. Vertical stacking mismatch boundary (SMB) would be formed at the edges of the BSF. Such SMB extends upward during growth and generates V-pits. These V-pits coalesce to form a trench loop encircling a central region[31-33]. The high-density trench defects are proven to significantly reduce internal quantum efficiency (IQE) in blue and green MQWs[25, 26, 34, 35]. The reduction in efficiency is attributed to the increased density of nonradiative recombination centers in MQWs, likely associated with SMBs[34]. Accordingly, trench defects are considered as defective structures leading to the degradation of LED efficiency. Various methods have been proposed to eliminate trench defects, such as increasing the MQWs growth temperature[23, 35, 36] and introducing hydrogen treatment during the growth of quantum barriers (QBs)[20, 21, 24].

Nonetheless, some studies indicate that the emission from regions inside trench loops redshifts and becomes more intense than the surrounding planar regions under certain circumstances[20, 37-39]. Some studies report that the wavelength redshift is attributed to the increased In incorporation, which may related to the strain relaxation effect induced by the encircling V-pits structure of trench defects[40, 41]. The reason for luminescence enhancement inside trench loops is not fully clarified yet. Massabuau et al. reported that those trench defects, emitting intense and redshift light, were typically accompanied by the BSF formation in the early growth stages of the MQWs[41, 42]. These underlying BSFs and SMBs do not significantly affect the upper QWs' luminescence[41, 42]. Recent research from the same group suggests that the strain relaxation effect within these trench defects plays a vital role in enhancing radiation recombination, as demonstrated by time-resolved cathodoluminescence (CL)[43]. However, the effect of defect-related recombination and Auger recombination on luminescence has not been sufficiently investigated. Besides, the wavelength redshift and luminescence enhancement are limited to small regions inside the trenches, while trench defects are still detrimental to the overall luminescence efficiency of the MQWs[34]. Therefore, proper modulation is necessary to apply the wavelength redshift and intensity enhancement effects of trench defects toward realizing high-efficiency red InGaN LEDs.

In this work, trench defects were modified as the luminescent sources to achieve high-efficiency red InGaN LEDs. Dual-color MQWs structures were grown, including green MQWs at the bottom and red MQWs at the top. By lowering the growth temperatures of green QWs and QBs, a high density of trench defects was induced in the green MQWs. Impressively, red MQWs grown on green MQWs with trench defects exhibited significant wavelength redshift and luminescence enhancement compared to those without trench defects. The underlying mechanisms of the wavelength redshift and luminescence enhancement were explored in detail. The emission microstructures of MQWs were studied by hyperspectral CL imaging and confocal photoluminescence (PL). These results indicate that trench defects can serve as effective luminescent structures in In-rich MQWs, boosting the efficiency of red InGaN LEDs.

## 2. Results and Discussion
### 2.1. Modulation of Trench Defects

**Figure 1** shows the schematic structure of samples regarding the modulation of trench defects. Three green MQWs samples (A1, B1, and C1) with different trench defect densities were obtained by adjusting the QW/QB growth temperatures, as shown in Figure 1a. The QW/QB growth temperatures of samples A1, B1, and C1 were 700 °C/820 °C, 675 °C/760 °C, and 630 °C/700 °C, respectively. Lowering the QW temperatures aims to increase the In-rich surfaces in QWs[29], whereas reducing the QB temperatures diminishes the Ga adatom mobility during QB growth[44]. The combined effect of these conditions significantly increases the trench defect density (Figure S1, Supporting Information). Although the growth temperatures for green MQWs varied, the In components were kept nearly consistent by adjusting the TMIn flux. Furthermore, three full-structure LED samples (A2, B2, and C2) with red MQWs and p-GaN layers were grown based on the above green MQWs, as shown in Figure 1b. The only difference among these samples was the trench defect densities in the green MQWs. The growth parameters of the red MQWs were identical. Red MQWs structures without p-GaN layers were also grown to investigate the relationship between surface morphology and emission properties.

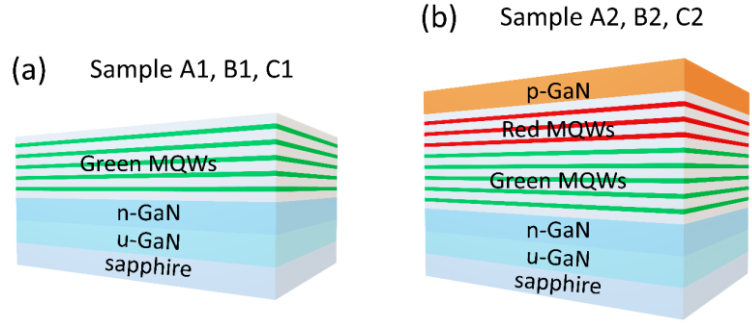

Figure 1. Schematic structures of (a) green MQWs samples A1, B1, and C1 with different trench defect densities. (b) The full-structure red LED samples A2, B2, and C2. The C serial samples show the highest trench defect densities in green MQWs.

**Figure 2** shows the surface morphologies of the green MQWs in samples A1, B1, and C1. The V-pit densities of samples A1, B1, and C1 are all around $(1.5\pm0.3)\times10^8$ cm$^{-2}$. No obvious trench defects appear on the surface of sample A1. As for samples B1 and C1, the trench defect densities are about $(5.5\pm0.5)\times10^8$ cm$^{-2}$ and $(4.5\pm0.3)\times10^9$ cm$^{-2}$, respectively. The trench defect density increases with decreasing QW and QB growth temperatures. The scanning electron microscopy (SEM) and atomic force microscopy (AFM) images demonstrate distinct morphological characteristics of trench defects. The SEM images show clearer 2D morphologies due to the little De Broglie wavelength of electrons. The AFM images show better 3D profiles, clearly displaying the pits and surface prominence. Trench defect morphologies can be classified by three characteristics: (i) the width of the trench, (ii) the area inside the trench, and (iii) the prominence above surroundings. Previous studies reported that wavelength redshift and luminescence enhancement effects in the MQWs inside trench loops are strongly correlated with the trench width, which directly relates to the location of BSF generation[41, 42]. Further, wavelength redshift and luminescence enhancement can be observed in the top QWs inside the trench only if the BSFs are generated in the bottom few QWs stack[41, 42]. A high density of trench defects has appeared on the surface of the green MQWs, as presented in Figure 2b-c and 2e-f. The appearance of trench loops indicates the inclusion of both BSFs and SMBs within the green MQWs.

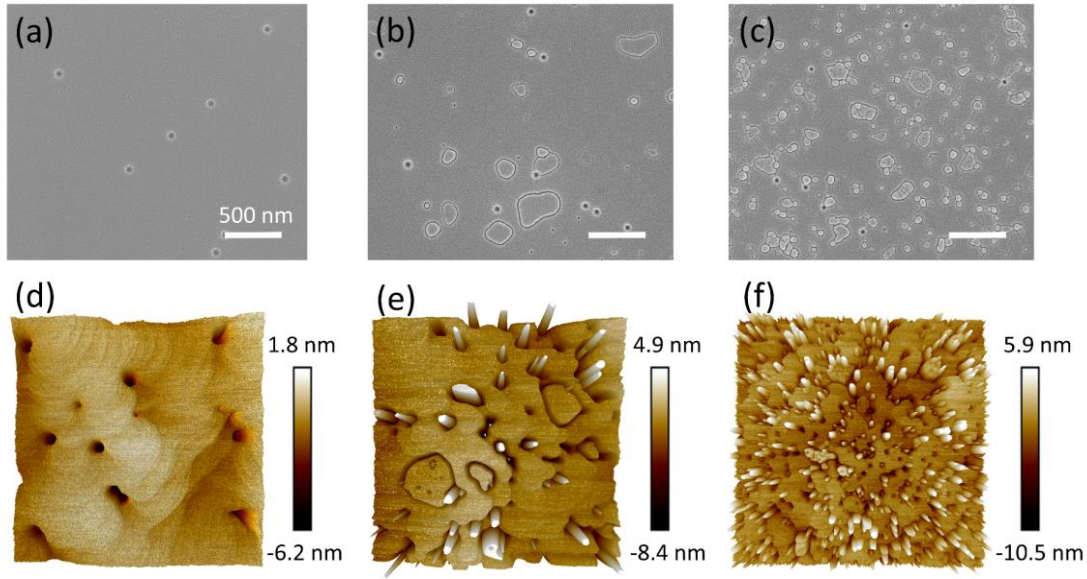

Figure 2. (a-c) SEM, (d-f) AFM images of green MQWs samples A1, B1, and C1. All scale bars denote 500 nm. The AFM images correspond to 3 μm square regions. The trench defect density increases with decreasing MQWs growth temperatures.

**2.2. Effect of Trench Defects on the Luminescence of MQWs**

**Figure 3**a shows the PL spectra of green MQWs samples A1, B1, and C1 under 405 nm laser excitation, with peak wavelengths around 510 nm. With the increasing density of trench defects, a substantial decrease in PL intensity of the green MQWs is observed. The PL intensity reduction is mainly related to defect structures, such as SMBs in trench defects and In-rich clusters in QWs [33, 45]. The corresponding SEM-CL images of green MQWs (Figure S2, Supporting Information) show that most of the trench defects in green MQWs exhibit dimmer luminescence in CL images, mainly associated with embedded defects such as SMBs[33].

Figure 3b-c presents PL spectra of the full-structure red LED samples A2, B2, and C2 under 405 nm and 532 nm laser excitation. Since the photon energy of the 532 nm laser lies between the bandgaps of the red and green QWs, it is feasible to excite the carriers in red MQWs selectively. Interestingly, although the luminescence from green MQWs weakens with increasing trench defect density, the PL intensities of the red MQWs gradually increase, accompanied by a significant redshift in peak wavelength. The peak wavelengths are 569, 615, and 653 nm for samples A2, B2, and C2. The PL intensities enhanced 3.1 and 13.6 times for samples B2 and C2 compared to sample A2. In addition, the fluorescence images of samples A2, B2, and C2 under 405 nm laser excitation further indicate that with an increase in trench defect density, the luminescence of the green MQWs drops dramatically while the luminescence of the red MQWs notably increases (see Figure S3, Supporting Information).

The temperature-dependent PL was used to probe the IQEs of the red MQWs. The IQE is considered as the ratio of the PL integral intensity at room temperature to that at 3 K. Figure 3d-e demonstrates the IQEs and corresponding peak wavelengths of red MQWs at various power densities of a 532 nm laser. The IQE values increase, and peak wavelengths blueshift with increasing laser power density for these samples. There are no IQE droops until the laser power density reaches $10^5$ W/cm$^2$. The IQEs of samples A2, B2, and C2 were measured as 2.7% at 539 nm, 6.9% at 582 nm, and 16.4% at 607 nm under a maximum excitation power density of $1.2 \times 10^5$ W/cm$^2$. It confirms that the red MQWs exhibit efficiency enhancement and wavelength redshift with increasing density of trench defects in green MQWs.

Figure 3f shows the normalized electroluminescence (EL) spectra of samples A2, B2, and C2 under a current of 40 mA. The EL peak wavelengths of samples A2, B2, and C2 are 547, 577, and 607 nm, respectively. It shows an approximately 60 nm redshift from green towards the red region with increasing trench defect density, similar to the redshift observed in the PL measurements. The single emission peaks for the dual-color MQWs are due to hole mobility limitations, and only the upper red MQWs contribute to the emissions. Generally, higher In components in MQWs are accompanied by lower radiative recombination efficiency due to poorer crystal quality and stronger polarization electric field. However, higher In component InGaN QWs show higher luminescence efficiency in this work. It indicates that introducing high-density trench defects in the underlying green MQWs can not only improve the In incorporation in the red MQWs, but also improve the radiative recombination efficiency. Thus, further investigations at the microscopic scale are necessary to clarify the mechanism behind such efficiency enhancement.

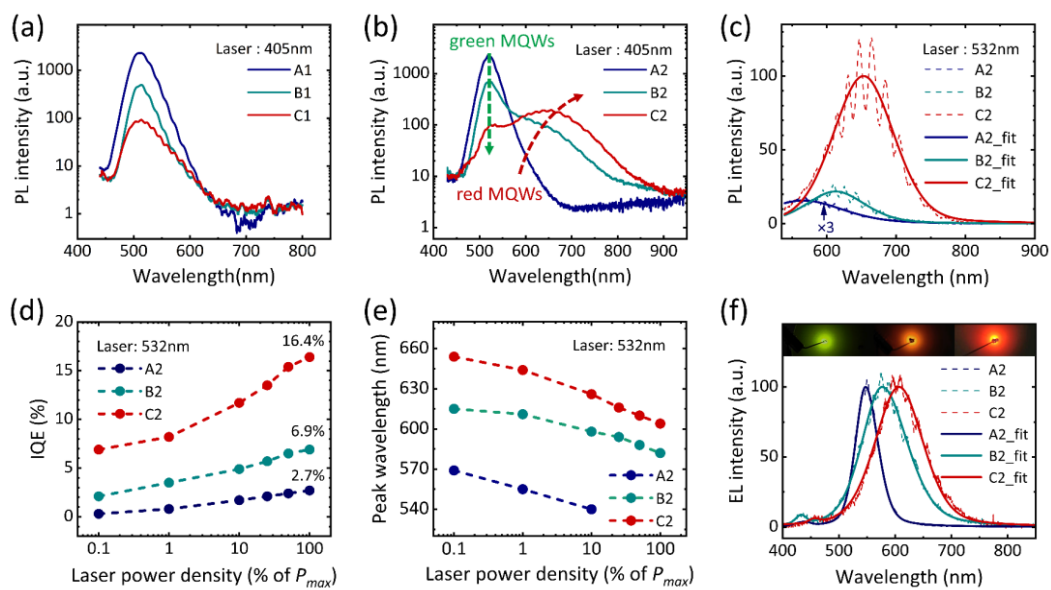

Figure 3. PL spectra of (a) the green MQWs samples A1, B1, and C1 under 405 nm laser excitation, and the full-structure red LED samples A2, B2, and C2 under (b) 405 nm and (c) 532 nm laser excitation. The dependencies of (d) IQEs and (e) peak wavelengths of samples A2, B2, and C2 on power density under 532 nm laser excitation, the $P_{max}$ corresponds to $1.2 \times 10^5$ W/cm$^2$. (f) EL spectra under 40 mA of samples A2, B2, and C2, the insets correspond to the EL images of samples A2, B2, and C2 from left to right. The dashed and solid lines in (c) and (f) correspond to the measured spectra and those obtained after Gaussian-Lorentzian fitting.

**2.3. Hyperspectral CL Imaging on Trench Defects**

Hyperspectral measurements were employed by CL mapping to explore the effect of trench defects on the efficiency enhancement of red MQWs at the nanoscale. The trench defects in sample C2 are too close to discriminate the emissions from different trenches, considering the carrier diffusion length (see Figure S4, Supporting Information). The MQWs-interrupted sample B2 was utilized to explore the relationship between surface morphology and emission properties of MQWs. **Figure 4**a-c shows the AFM and SEM images of red MQWs, localized to the same area by special markers. There are V-pits and trench defects of varying morphologies on the surface of the red MQWs. Figure 4d-e shows the images of the CL intensity maps obtained by integrating the green and red spectra after split-peak fitting, representing the microscopic luminescence images of the green and red MQWs, respectively. Figure 4f shows the CL peak wavelength map of red emissions.

In Figure 4d, the green emission predominantly originates from the area outside the trench. The emission intensity inside the trench is much lower, mainly related to the SMBs[33]. Besides, luminescence enhancement can be observed at the edges of the trench defects and V-pits, attributed to the increase in light extraction efficiency due to the surface structures. In Figure 4e, the average integral intensity of the red emission within trench loops is enhanced by about 2.5 times compared to the surrounding region. It indicates that introducing trench defects effectively promotes the radiative recombination efficiency of the red MQWs within the trench loops. Moreover, an average wavelength redshift of about 29 nm is observed within the trench loops compared to the surrounding region, as shown in Figure 4f. It is mainly related to the higher In component of red MQWs inside the trench[41]. The scanning transmission electron microscopy combined with energy-dispersive X-ray spectroscopy (STEM-EDX) profiles clearly show the higher In components of red MQWs inside the trench than outside (Figure S5, Supporting Information). The luminescence enhancement and wavelength redshift within the trench defects have also been observed in other studies[37, 41].

Nevertheless, a comparison of AFM/SEM and CL images reveals that not all trench defects exhibit luminescence enhancement and wavelength redshift. The correlation between surface morphology and emission properties in red MQWs has been investigated in detail (see Figure S6-S9, Supporting Information). It is observed that the emission properties of red MQWs inside the trench are mainly related to the width of the trench and the prominence above surroundings. Notably, luminescence enhancement and wavelength redshift only occur inside trench defects with wide grooves. For trench defects with wide grooves, an excessive prominence generally corresponds to a smaller enhancement magnitude. It is known that the sidewall facets of the trench are determined under certain MQWs growth conditions, indicating that the wider trench means the deeper trench. The BSFs of deep trench defects generally form in the underlying green MQWs. As a result, the region inside trench defects intentionally introduced in the underlying green MQWs serve as effective luminescent centers in the upper red MQWs.

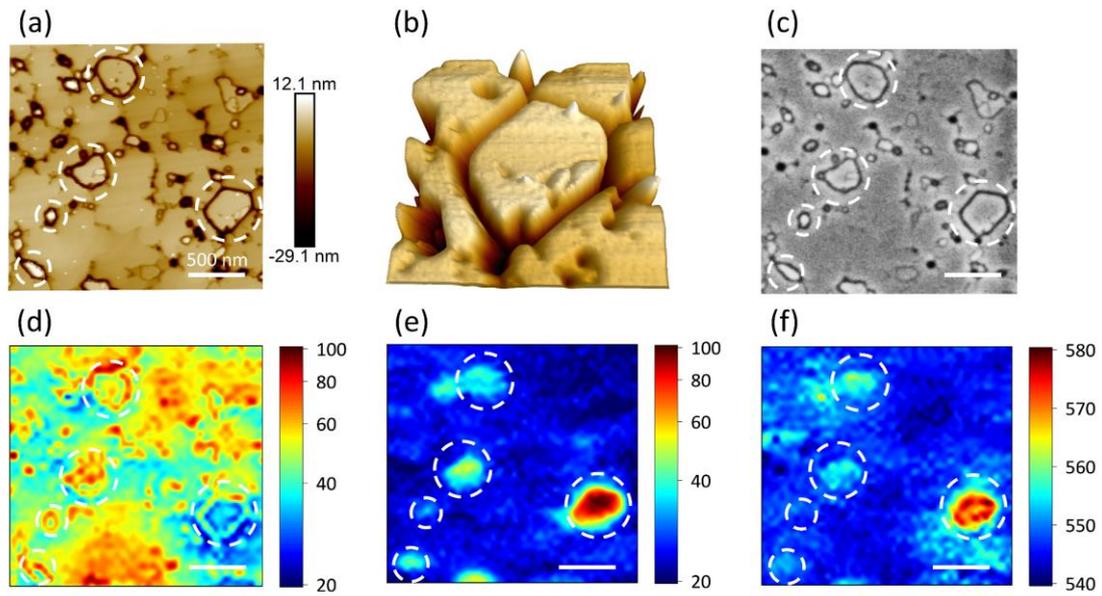

Figure 4. (a) AFM image of the MQWs-interrupted sample B2. (b) Enlarged view of trench defects in the rightmost circular dashed frame in (a). (c) SEM image of the same region in (a). The corresponding CL integral intensity maps of (d) the green MQWs and (e) the red MQWs. (f) The CL peak wavelength map of the red MQWs. All scale bars denote 500 nm.

The deep trench defects contribute to CL intensity enhancement and wavelength redshift locally inside the trench, as mentioned above. The overall impacts of trench defects on the red MQWs remain unclear. Here, large-scale hyperspectral CL imaging was performed on the full-structure red LED samples. **Figure 5** demonstrates the CL intensity maps of green and red MQWs for samples A2, B2, and C2. In Figure 5a-c, the green MQW exhibits continuous flocculent-like luminescence. The green floccules correspond to the continuous regions apart from the defects in QWs[33, 45, 46]. With decreasing the growth temperatures of green MQWs, the effective luminescence area shrinks and the dark area expands, which coincides with the decrease of PL intensity in Figure 3a.

Figure 5d shows the CL intensity map of the red MQWs in sample A2. It also shows flocculent-like luminescence, similar to green MQWs in Figure 5a. The formation of a flocculent luminescence may be due to the low In incorporation and little strain relaxation. Nonetheless, the luminescence efficiency of red MQWs in sample A2 remains relatively low, as shown in Figure 3d. In fact, the low growth temperatures of red MQWs result in the new formation of shallow trench defects with density of $(5.7\pm0.5)\times10^8$ cm$^{-2}$ (see Figure S4a, Supporting Information). The SMBs and BSFs in these shallow trench defects significantly reduces the radiative recombination efficiency, as confirmed by SEM-CL results (Figure S10, Supporting Information).

Figure 5e-f shows the CL intensity maps of the red MQWs in samples B2 and C2. It exhibits dot-like luminescence, which originates from the strong emission inside trench defects, as confirmed by Figure 4. With increasing density of trench defects in the underlying green MQWs, the density of dot-like luminescence in the red MQWs increases, and the wavelength exhibits a redshift, which coincides with the PL and EL results in Figure 3. The SEM images of the red MQWs (Figure S4b-c, Supporting Information) reveal that the trench defect densities in red MQWs of samples B2 and C2 are $(1.2\pm0.3)\times10^9$ cm$^{-2}$ and $(5.3\pm0.3)\times10^9$ cm$^{-2}$, respectively. The trench defect densities in the red MQWs are substantially higher than that in the underlying green MQWs, implying that large numbers of shallow trench defects are newly generated in the red MQWs. However, more shallow trench defects do not result in weaker luminescence for samples B2 and C2. In Figure 5e-f, the emission within the deep trench defects remains unaffected by newly generated shallow trench defects, mainly owing to the blocking effect of the trench on the carriers. It suggests that those deep trench defects contributing to the luminescence exhibit a spatial isolation effect for external defects. Based on the above results, the efficiency enhancement with increasing trench defect density mainly originates from the strong emission inside deep trench defects.

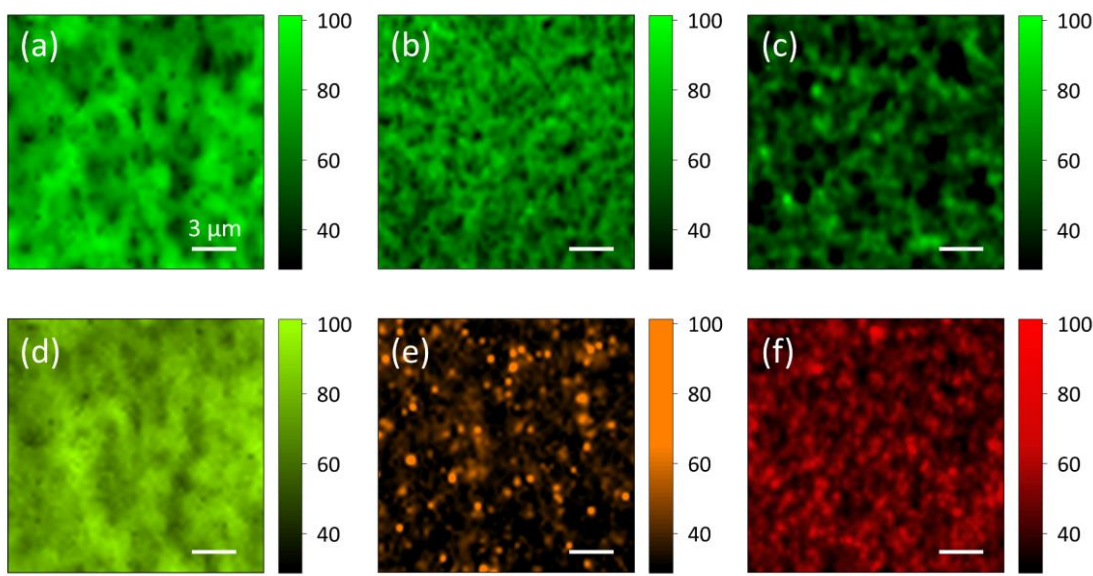

Figure 5. The integrated CL intensity maps of (a-c) green MQWs and (d-f) red MQWs of the full-structure LED samples A2, B2, and C2. Images for green and red emissions correspond to the same region in each sample. All scale bars denote 3 μm.

## 2.4. Confocal PL Mapping with Different Injections on Trench Defects

The above CL results provide the nanoscale luminescence images of the MQWs. Nevertheless, the actual luminescence image of red MQWs under electrical injection does not exactly coincide with the CL images due to CL's high energy injection[46, 47]. Confocal PL mapping at variable power densities was performed on sample B2 since the PL allows excitation at lower injections that approach the electrical injection[48]. Confocal PL can obtain luminescence images with a spatial resolution of 300 nm. **Figure 6**a-d shows the PL integral intensity distribution in the same region as the laser power density changing from $3.1\times10^3$ W/cm$^2$ to $3.1\times10^6$ W/cm$^2$. The PL intensity maps of sample B2 exhibit dot-like luminescence, consistent with the CL results in Figure 4-

5. Thus, the bright spots in PL mapping mainly come from the region inside trenches. In Figure 6d, the regions inside and outside trenches could both emit effectively, with the region inside trenches exhibiting a stronger luminescence, similar to the CL results. At high injection, the carriers exhibit short lifetimes and small diffusion lengths, and the polarization field is partially shielded. The defects can be filled and saturated, leading to more emissions from the defective regions outside trench loops[26]. At low injections, only the red MQWs inside trenches can emit light efficiently as shown in Figure 6a. The carrier lifetime becomes longer and the diffusion length increases at low injections. Such long lifetime and diffusion length of carriers will make the recombination predominantly at the defects and/or potential minima. Meanwhile, the polarization field significantly affects the luminescence. In this case, the emission from the surrounding region becomes substantially dimmer, with the primary luminescence coming from the region inside trenches.

The spectra of 35 points in bright spots and 350 points in dark areas were selected with different laser power densities. Figure 6e-f presents the dependencies of the peak wavelength and integrated intensity of their average spectra on the laser power density. With the laser power density increasing from $3.1 \times 10^3$ to $3.1 \times 10^6$ W/cm$^2$, the peak wavelengths of the red QWs outside trenches shift from 608 nm to 551 nm with a blueshift of 57 nm. Further X-ray diffraction reciprocal space mapping (XRD RSM) results show that the MQWs outside trenches are in full strain with the underlying GaN layer (see Figure S11, Supporting Information). In comparison, the peak wavelengths of the red QWs inside trenches shift from 599 nm to 563 nm, exhibiting a blueshift of 36 nm. The peak wavelength inside trenches shows a smaller blueshift with increased laser power density. It indicates that the polarization electric field of the red QWs inside trenches is weaker than that outside, attributing to strain relaxation inside trenches.

In Figure 6f, the PL integral intensity inside trenches is about 24.7 times higher than outside the trench at the laser power density of $3.1 \times 10^3$ W/cm$^2$. It decreases to 2.9 times at the power density of $3.1 \times 10^6$ W/cm$^2$. Based on the ABC model, the relationship between the laser power $P_{laser}$ and the integral intensity $I_{PL}$ can be expressed as[49]:

$$P_{laser} = \frac{A}{x\sqrt{yB}} \sqrt{I_{PL}} + \frac{1}{xy} I_{PL} + \frac{C}{x(yB)^{3/2}} I_{PL}^{3/2} \qquad (1)$$

where A, B, and C are the respective recombination coefficients of Shockley-Read-Hall (SRH) recombination, radiative recombination, and Auger (and/or possibly carrier overflow) recombination. The $x$ is the coefficient between laser power and total carrier generation rate. The $y$ is a constant determined by the volume of the excited active region and the total collection efficiency of luminescence.

The superlinearity, linearity, and sublinearity of $I_{PL}$-$P_{laser}$ indicate the dominance of SRH recombination, radiative recombination, and Auger recombination in the active layer[49, 50]. The slope $k$ in log-log coordinates can represent the superlinear ($k > 1$), linear ($k = 1$), and sublinear ($k < 1$) relationships between $I_{PL}$ and $P_{laser}$. In Figure 6f, the curves show linear inside trenches and superlinear outside the trench in the range from 0.1% to 1% of $P_{max}$. Under low excitation, carriers inside trenches are mainly involved in radiative recombination and exclude the impact of defects, while the carriers outside trenches are severely affected by SRH recombination related to defects.

As mentioned above, more severe QCSE in the red MQWs outside trenches results in weaker luminescence than inside. These factors cause the low luminescence efficiency outside the trench at low injection. From 10% to 100% of $P_{max}$, the curves show sublinear both inside and outside trenches. It shows a lower slope in the $I_{PL}$-$P_{laser}$ curves inside trenches. When the laser power density increases, excessive non-equilibrium photo-generated carriers can shield the polarization field, thus weakening the QCSE in MQWs[9, 11] outside the trench. The SRH recombination will tend to saturate and have a weak effect on luminescence[51]. Besides, a higher carrier concentration occurs inside trenches since the carriers are confined within the trench. Higher carrier concentrations inside trenches will result in more Auger recombination, leading to more nearly sublinear $I_{PL}$-$P_{laser}$ curves. Accordingly, these factors combine to reduce the PL intensity gap between the regions inside and outside trenches at high injection.

To clarify the effect of polarization, SRH recombination, and Auger recombination on the luminescence, simulations were performed using Crosslight APSYS software[52] (see Figure S12, Supporting Information). In Figure 6g, the energy band inside trenches is narrower than outside due to higher In components. Meanwhile, the red MQWs outside trenches suffer from more severe energy band bending than inside. Figure 6h indicates that the polarization effect of red MQWs outside trench defects is the primary factor contributing to weaker luminescence than that inside trench defects at high injection. Nevertheless, SRH recombination is the predominant reason for the drastic decrease in the luminescence outside the trench defects at low injection. In addition, the carriers inside the trench defects suffer from more Auger recombination. At higher injection levels, the gap in luminescence between the inside and outside of trench defects can be predicted to decrease further. Consequently, the red MQWs inside trenches exhibit a weaker polarization effect, less SRH recombination, and more Auger recombination compared to those outside. Especially at low injection, the red MQWs inside trenches can still emit efficiently without being overly affected by defects, while the red MQWs outside emit light weakly.

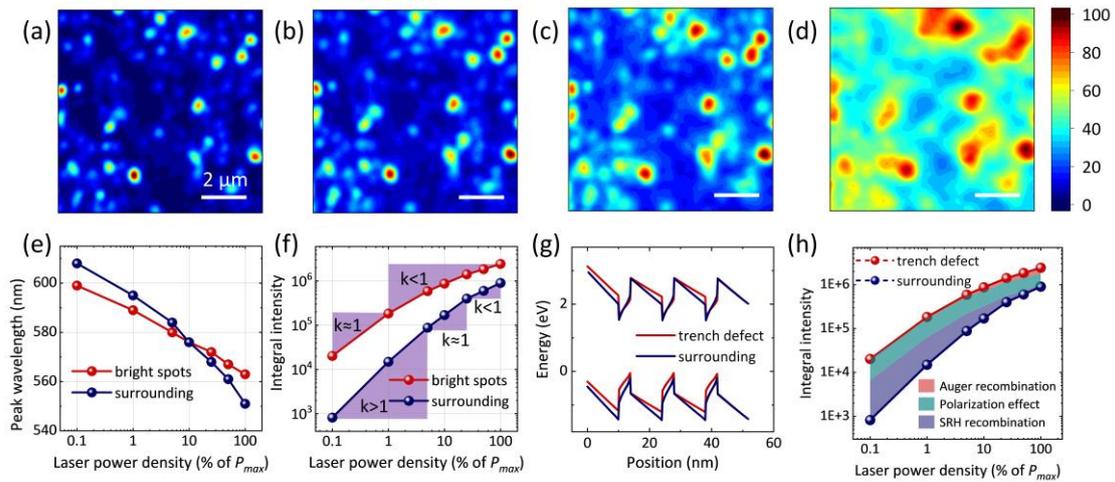

Figure 6. Integrated PL intensity maps of red MQWs in sample B2 under 532nm laser excitation at the power density of (a) 3.1×10³, (b) 3.1×10⁴, (c) 3.1×10⁵, and (d) 3.1×10⁶

W/cm$^2$. All scale bars denote 2 μm. Dependencies of (e) peak wavelength, (f) integral intensity, on the laser power density in the regions inside and outside trench defects. The slope of the line is labeled in (f). (g) Energy band diagrams of the MQWs at the laser power density of 3.1×10$^6$ W/cm$^2$ inside and outside trench defects. (h) Dependencies of PL integral intensity of the MQWs inside and outside trench defects on the laser power. The $P_{max}$ corresponds to 3.1×10$^6$ W/cm$^2$.

## 2.5. Correlation between Microstructures and Emission Properties of Trench Defects

To investigate the effect of trench defects on red emissions, dark-field transmission electron microscopy (TEM), high-angle annular dark field scanning transmission electron microscopy (STEM-HAADF), and STEM-EDX were applied. **Figure 7**a-c shows the cross-sectional dark-field TEM, STEM-HAADF, and STEM-EDX images of a trench defect. A BSF is observed at the second GaN/InGaN interface of green MQWs near n-GaN. A SMB appears at the edge of the BSF. Notably, the V-pits arise in the initial stages of the SMB formation. Subsequently, the width and depth of V-pits increase as the SMB extends upward. In Figure 7b-c, the QWs at the sidewalls of V-pits exhibit narrower widths and lower In components compared to those in the planar regions. The growth up of V-pits in Figure 7b means the sidewall facets show a slower growth rate than that of the c-plane, which corresponds to the longer side length of upper V-pits. Thus, the carriers in the planar region are difficult to diffuse into the SMBs due to the blocking effect of the wider bandgaps and longer distances of the sidewall QWs, similar to the mechanism of V-pits induced by threading dislocations (TDs)[16, 53]. Notably, the vertices of V-pits induced by TDs are generally located underneath the MQWs, while the vertices of V-pits in trench defects are located inside the MQWs. For V-pits induced by TDs, the QWs in the planar region remain about tens of nanometers away from the dislocation inside V-pits. However, for V-pits in trench defects shown in Figure 7b, the planar QWs at the bottom are only a few nanometers away from the SMBs, which may expose the carriers in the QWs to severe defect-related recombination. When the V-pits expand upward to larger widths and depths, the carriers in the topmost planar QWs can be shielded from the SMBs. It explains why the luminescence enhancement of red MQWs occurs only inside deep trenches.

Figure 7d-e demonstrates the Bragg-filtered and high-magnification STEM-HAADF images in the defective region. In Figure 7d, a misalignment of atoms within the trench defects with the underlying atoms occurs, and a lateral dark contrast confirms the existence of the BSF. A vertical dark contrast of the SMB can be clearly observed at the edge of the BSF. In Figure 7e, the schematic arrangement of the atoms in different colors shows the stacking order of the atoms. At the interface of the BSF, the arrangement order of the atoms changes from ABABAB to BCBCBC, as delineated by the white dotted line. It indicates that the BSF is of type I$_1$. Meanwhile, the atoms on the sides of the SMB are arranged in a different order, changing from ABABAB to BCBCBC. Thus, the BSF leads to the misalignment of atoms inside trench defects, resulting in the formation of massive SMBs at the edge of the BSF.

Figure 7f-g shows the STEM-HAADF and Bragg-filtered images of the red MQWs

inside and outside the trench, corresponding to the regions within the red dashed rectangles in Figure 7b. The red MQWs inside the trench exhibit superior lattice ordering compared to that outside the trench, indicating a lower defect density. It coincides with our observation in confocal PL results that red MQWs inside the trench suffer from less SRH recombination. Considering the growth process from the perspective of In incorporation, the trench loop consisting of V-pits provides a significant strain relaxation effect, allowing efficient In incorporation without introducing excessive defects inside the trench. By contrast, less strain relaxation in the surrounding region results in defect formation for heavy In incorporation[54, 55].

Figure 7h illustrates a schematic of the sectional structure of a trench defect. Despite the higher In component, the red MQWs inside the trench exhibit stronger emissions than those in the surrounding region. Some explanations for the strong emission of the red MQWs inside the trench can be given on the basis of the experimental results. Firstly, the shielding effect of trench defects suppresses nonradiative recombination in their internal red MQWs. The topmost red MQWs inside the trench spatially isolate BSFs and SMBs, as depicted in Figure 7a-b. Meanwhile, the blocking of the trench prevents the red MQWs inside the trench from being affected by defects in the surrounding region, similar to defect shielding by nanorods[56]. As a result, the red MQWs inside the trench would shield the defects in both the growth direction and in-plane dimensions, effectively reducing their nonradiative recombination. Secondly, the trench defects can relax the compressive strain locally, effectively weakening the polarization electric field of red MQWs inside the trench loops, as demonstrated by confocal PL results. The weakening of the polarization electric field will alleviate the QCSE, facilitating the radiative recombination. Thirdly, the red MQWs inside the trench loops exhibit a lower defect density than that in the surrounding region. Local strain relaxation inside the trench reduces the strain energy required for In-N bond, promoting In incorporation rather than relieving strain by generating excess defects[14, 54, 55]. This effect significantly lowers the defect density and reduces SRH recombination inside the trench, as demonstrated by confocal PL and STEM-HAADF results. It is suggested that the stronger luminescence inside the trench defects compared to the surrounding region is mainly attributed to the shielding effect, a weaker polarization electric field, and a lower defect density.

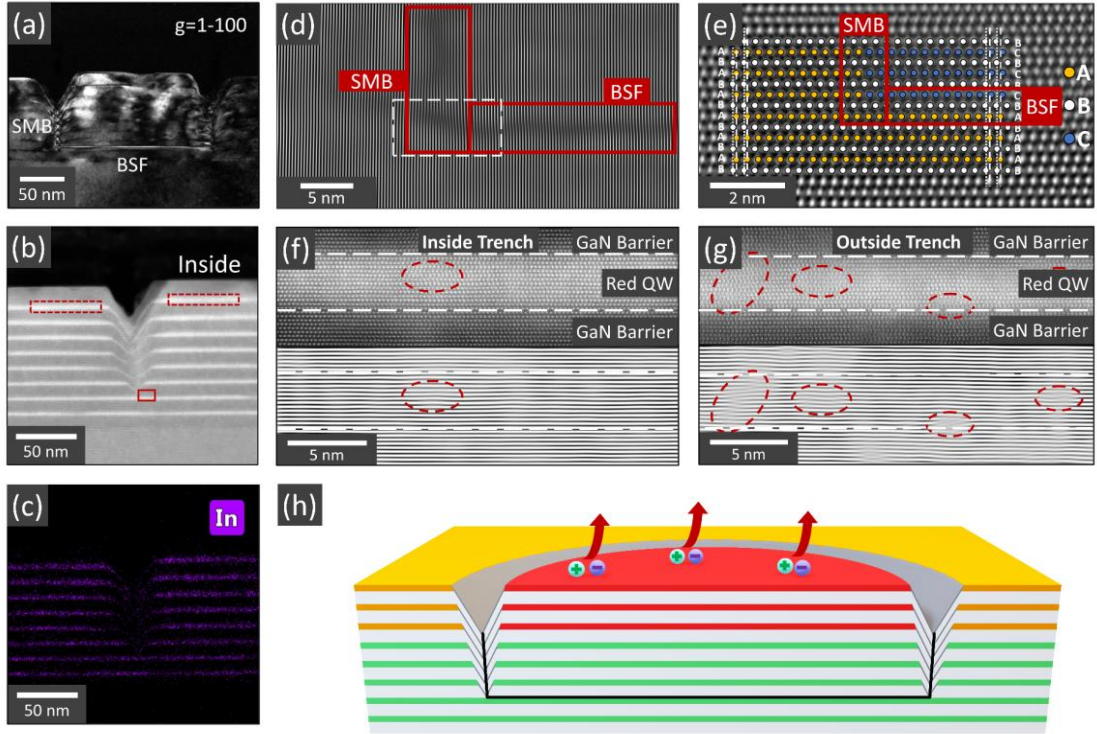

Figure 7. (a) Cross-sectional dark-field TEM image of a trench defect (zone axis 11-20) with **g** (1-100). (b) STEM-HAADF image of the trench defect. (c) Corresponding In atom EDX mapping. (d) Bragg-filtered image comprising (1-100) lattice fringes image of the region in the red solid rectangle in (b), and the solid red rectangle indicates the location of the BSF and SMB. (e) Magnified STEM-HAADF image of the region at the BSF and SMB junction in the white dashed rectangle in (d). Magnified STEM-HAADF and corresponding Bragg-filtered images of a red QW (f) inside the trench and (g) outside the trench, correspond to the regions within the red dashed rectangle in (b), respectively. The white dashed lines delineate the region of QWs, while the red dashed elliptical boxes indicate the location where the lattice mismatch occurs. (h) Schematic of the structural properties of trench defects.

In this work, trench defects are utilized effectively to enhance the luminescence efficiency of red LEDs. Indeed, trench defects can be easily introduced in MQWs by various methods[22, 23], typically at low growth temperatures. From the perspective of growth kinetics, the Ga adatom diffusion length is dramatically shortened as the growth temperatures decrease. Once the diffusion length shortens below the average distance between the binding sites, surface roughening occurs. The Ga adatoms may be trapped in the fcc site and not diffuse to the hcp site, leading to the formation of BSFs[29, 44]. Nevertheless, the formation locations of BSFs are inherently random, while concurrently inducing trench defects with diverse morphologies. The varying prominences inside the trench defects represent a difference in the growth rate or lattice parameters inside and outside the trench under the same conditions. In order to maximize the luminescence enhancement effect of trench defects, precise modulation should be performed to raise the proportion of deep trench defects with low prominence as much as possible, or to make the trench defect density high enough to approach fully closed-packed structures. Further modulation can potentially be achieved by optimizing the methods of introducing trench defects. Currently, there is a lack of effective methods for modulating trench defects with specific morphology. The formation mechanism of trench defects with different morphologies needs to be further explored.

## 3. Conclusions

Trench defects were innovatively utilized to improve the luminescence efficiency of red InGaN LEDs. By appropriately modulating the trench defects, the red MQWs show significant wavelength redshift and luminescence enhancement. The wavelength redshift is attributed to the increased In component due to the strain relaxation effect of trench defects. The luminescence enhancement originates from the strong emission inside trench defects. Especially at low injection, the red MQWs inside trenches serve as the primary contributor to emission. The stronger luminescence inside trenches than outside is mainly attributed to the shielding effect, weaker polarization electric field, and lower defect density. The carriers in the topmost QWs inside trenches suffer from less SRH recombination, weaker QCSE, and more Auger recombination compared to those outside. Overall, this research provides a new perception of trench defects and explores new avenues for achieving high-efficiency red InGaN LEDs.

## 4. Experimental Section

*Growth Method:* All epitaxial layers were grown on c-plane sapphire substrates in an Aixtron Crius I 31×2 inch close-coupled showerhead MOVPE system in this study. The specific structures of the green MQWs samples (A1, B1, C1) consist of a 1 μm unintentionally doped (UID) GaN layer, 2 μm Si-doped n-GaN layer, and five green MQWs composed of 2.5/3/7 nm of $In_xGa_{1-x}N$ well/GaN cap/UID GaN barrier layers. The UID GaN and n-GaN layers were grown with TMGa and $NH_3$ under $H_2$ ambient, while the MQWs layers were grown with TEGa, TMIn, and $NH_3$ under $N_2$ ambient. The QW/QB temperatures of samples A1, B1, and C1 were 700 °C/820 °C, 665 °C/760 °C, and 630 °C/700 °C, respectively. By adjusting the ratios of TMIn flux, the In components in green MQWs were kept nearly consistent. The corresponding TMIn/TEGa ratios were 700 sccm/140 sccm, 300 sccm/140 sccm, and 220 sccm/140 sccm, respectively. The GaN cap layers were grown at the same temperatures as the QWs. For full-structure red LED samples (A2, B2, C2), three pairs of red MQWs are composed of 3.5/3/7 nm of $In_yGa_{1-y}N$ well/GaN cap/UID GaN barrier layers, with the QW layers and GaN cap layers being grown at 665°C and the QB layers at 760 °C. By employing a larger TMIn/TEGa ratio, the red MQWs correspond to a higher In component compared to the green MQWs. The p-GaN layer consisted of a p+-GaN layer grown at 840 °C, a p-GaN layer grown at 935 °C, and a p-InGaN contact layer grown at 700 °C. The growth temperatures mentioned above are epiwafer surface temperatures instead of heater temperatures.

*Measurements:* SEM (FEI Navo NanoSEM 430) and AFM (Bruker Dimension Icon) were employed to investigate the surface morphology of MQWs. In order to obtain the exact morphology of the trench defects, ultra-fine AFM probes (Bruker

FIB3D2-100A) with a high aspect ratio were utilized. SEM (ThermoFisher Quattro ESEM) combined with hyperspectral CL (Goldenscope Rainbow) imaging was performed to obtain luminescence spectra for different microstructures of MQWs. The spatial resolution of CL is around 50 nm due to the diffusion length of high-energy electrons. Confocal PL mapping was accomplished on a confocal micro-Raman spectrometer system (Horiba LabRAM HR Evolution) with a 532 nm laser and a 150 lines per mm grating blazed at 500 nm. The spatial resolution of the confocal microscope is about 300 nm at 100X object. Fluorescence Microscope (Olympus BX51WI) with a 405 nm laser was utilized to observe the luminescence images of MQWs directly. Cross-sectional TEM specimens regarding trench defects were prepared on focused ion beam (FIB)-SEM (ThermoFisher Helios G4 UX). A marking method of electron beam carbon coating was employed to locate specific trench defects. Dark-field TEM (FEI Tecnai F20) was applied to identify the location of BSFs and SMBs. STEM-HAADF and STEM-EDX (FEI Titan Cubed Themis G2 300) were applied to obtain the atomic arrangement and elemental distribution around trench defects.

**Acknowledgements**
The authors gratefully acknowledge the National Key R&D Program of China (No. 2021YFB3600100, No. 2020YFC2008200), the National Natural Science Foundation of China (No. 62174004, No. 61927806, No. 81670870), the Basic and Applied Basic Research Foundation of Guangdong Province (No. 2020B1515120020), the Beijing-Tianjin-Hebei Special Project (No. J200014). The authors acknowledge Electron Microscopy Laboratory of Peking University, China, for the use of ThermoFisher Helios G4 UX, FEI Tecnai F20 TEM, and FEI Titan Cubed Themis G2 300 TEM.

**Conflict of Interest**
The authors declare no conflict of interest.

**Data Availability Statement**
The data that support the findings of this study are available from the corresponding author upon reasonable request.

**Keywords**
trench defects, V-pits, InGaN, red LEDs.

**References**
[1] H. E. Lee, D. Lee, T.-I. Lee, J. H. Shin, G.-M. Choi, C. Kim, S. H. Lee, J. H. Lee, Y. H. Kim, S.-M. Kang, S. H. Park, I.-S. Kang, T.-S. Kim, B.-S. Bae, K. J. Lee, *Nano Energy* **2019**, *55*, 454.
[2] J. Xiong, E. L. Hsiang, Z. He, T. Zhan, S. T. Wu, *Light Sci. Appl.* **2021**, *10*, 216.
[3] J. E. Ryu, S. Park, Y. Park, S. W. Ryu, K. Hwang, H. W. Jang, *Adv. Mater.* **2023**, 2204947.
[4] Z. Chen, B. Sheng, F. Liu, S. Liu, D. Li, Z. Yuan, T. Wang, X. Rong, J. Huang, J. Qiu, W. Liang, C. Zhao, L. Yan, J. Hu, S. Guo, W. Ge, B. Shen, X. Wang, *Adv. Funct. Mater.* **2023**, *33*, 2300042.
[5] D. Iida, K. Ohkawa, *Semicond. Sci. Technol.* **2021**, *37*, 013001.

[6] A. Koukitu, N. Takahashi, T. Taki, H. Seki, *J. Cryst. Growth* **1997**, *170*, 306.

[7] D. Cherns, S. J. Henley, F. A. Ponce, *Appl. Phys. Lett.* **2001**, *78*, 2691.

[8] D. D. Koleske, A. E. Wickenden, R. L. Henry, M. E. Twigg, *J. Cryst. Growth* **2002**, *242*, 55.

[9] V. Fiorentini, F. Bernardini, F. Della Sala, A. Di Carlo, P. Lugli, *Phys. Rev. B* **1999**, *60*, 8849.

[10] M. E. Aumer, S. F. LeBoeuf, S. M. Bedair, M. Smith, J. Y. Lin, H. X. Jiang, *Appl. Phys. Lett.* **2000**, *77*, 821.

[11] O. Ambacher, J. Majewski, C. Miskys, A. Link, M. Hermann, M. Eickhoff, M. Stutzmann, F. Bernardini, V. Fiorentini, V. Tilak, B. Schaff, L. F. Eastman, *J. Phys.: Condens. Matter* **2002**, *14*, 3399.

[12] G. B. Stringfellow, *J. Appl. Phys.* **1972**, *43*, 3455.

[13] F. C. LarchÉ, J. W. Cahn, in *Fundamental Contributions to the Continuum Theory of Evolving Phase Interfaces in Solids*, (Eds: Ball, J. M.,Kinderlehrer, D.,Podio-Guidugli, P.,Slemrod, M.), Springer Berlin Heidelberg, Berlin, Heidelberg 1999, 120.

[14] G. B. Stringfellow, *J. Cryst. Growth* **2010**, *312*, 735.

[15] T. L. Song, *J. Appl. Phys.* **2005**, *98*, 084906.

[16] S. Zhang, J. Zhang, J. Gao, X. Wang, C. Zheng, M. Zhang, X. Wu, L. Xu, J. Ding, Z. Quan, F. Jiang, *Photonics Res.* **2020**, *8*, 1671.

[17] F. Jiang, J. Zhang, L. Xu, J. Ding, G. Wang, X. Wu, X. Wang, C. Mo, Z. Quan, X. Guo, C. Zheng, S. Pan, J. Liu, *Photonics Res.* **2019**, *7*, 144.

[18] B. Damilano, N. Grandjean, S. Dalmasso, J. Massies, *Appl. Phys. Lett.* **1999**, *75*, 3751.

[19] L. Wang, L. Wang, J. Yu, Z. Hao, Y. Luo, C. Sun, Y. Han, B. Xiong, J. Wang, H. Li, *ACS Appl. Mater. Interfaces* **2018**, *11*, 1228.

[20] D. I. Florescu, S. M. Ting, J. C. Ramer, D. S. Lee, V. N. Merai, A. Parkeh, D. Lu, E. A. Armour, L. Chernyak, *Appl. Phys. Lett.* **2003**, *83*, 33.

[21] S. M. Ting, J. C. Ramer, D. I. Florescu, V. N. Merai, B. E. Albert, A. Parekh, D. S. Lee, D. Lu, D. V. Christini, L. Liu, E. A. Armour, *J. Appl. Phys.* **2003**, *94*, 1461.

[22] T. Suzuki, M. Kaga, K. Naniwae, T. Kitano, K. Hirano, T. Takeuchi, S. Kamiyama, M. Iwaya, I. Akasaki, *Jpn. J. Appl. Phys.* **2013**, *52*, 08jb27.

[23] F. C. P. Massabuau, A. Le Fol, S. K. Pamenter, F. Oehler, M. J. Kappers, C. J. Humphreys, R. A. Oliver, *Phys. Status. Solidi. (a)* **2014**, *211*, 740.

[24] F. Massabuau, M. Kappers, C. Humphreys, R. Oliver, *Phys. Status. Solidi. (b)* **2017**, *254*, 1600666.

[25] A. Tian, J. Liu, L. Zhang, Z. Li, M. Ikeda, S. Zhang, D. Li, P. Wen, F. Zhang, Y. Cheng, X. Fan, H. Yang, *Opt. Express* **2017**, *25*, 415.

[26] A. Tian, J. Liu, R. Zhou, L. Zhang, S. Huang, W. Zhou, M. Ikeda, S. Zhang, D. Li, L. Jiang, H. Lin, H. Yang, *Appl. Phys. Express* **2019**, *12*, 064007.

[27] D. Iida, Z. Zhuang, P. Kirilenko, M. Velazquez-Rizo, M. A. Najmi, K. Ohkawa, *Appl. Phys. Lett.* **2020**, *116*, 162101.

[28] T. Tange, T. Matsukata, T. Sannomiya, *Appl. Phys. Express* **2020**, *13*, 062004.

[29] Z. Shi, A. Tian, X. Sun, X. Li, H. Zang, X. Su, H. Lin, P. Xu, H. Yang, J. Liu, D. Li, *J. Appl. Phys.* **2023**, *133*, 123103.

[30] K. Prabakaran, R. Ramesh, P. Arivazhagan, M. Jayasakthi, S. Sanjay, S. Surender, S. Pradeep, M. Balaji, K. Baskar, *J. Alloys Compd.* **2019**, *811*, 151803.

[31] H. K. Cho, J. Y. Lee, C. S. Kim, G. M. Yang, N. Sharma, C. Humphreys, *J. Cryst. Growth* **2001**, *231*, 466.

[32] H. K. Cho, J. Y. Lee, G. M. Yang, C. S. Kim, *Appl. Phys. Lett.* **2001**, *79*, 215.

[33] F. C. P. Massabuau, S. L. Sahonta, L. Trinh-Xuan, S. Rhode, T. J. Puchtler, M. J. Kappers, C. J. Humphreys, R. A. Oliver, *Appl. Phys. Lett.* **2012,** *101*, 212107.

[34] F. C. P. Massabuau, M. J. Davies, F. Oehler, S. K. Pamenter, E. J. Thrush, M. J. Kappers, A. Kovács, T. Williams, M. A. Hopkins, C. J. Humphreys, P. Dawson, R. E. Dunin-Borkowski, J. Etheridge, D. W. E. Allsopp, R. A. Oliver, *Appl. Phys. Lett.* **2014,** *105*, 112110.

[35] X. Wang, F. Liang, D. Zhao, D. Jiang, Z. Liu, J. Zhu, J. Yang, W. Wang, *J. Alloys Compd.* **2019,** *790*, 197.

[36] J. Smalc-Koziorowska, E. Grzanka, R. Czernecki, D. Schiavon, M. Leszczyński, *Appl. Phys. Lett.* **2015,** *106*, 101905.

[37] J. Bruckbauer, P. R. Edwards, T. Wang, R. W. Martin, *Appl. Phys. Lett.* **2011,** *98*, 141908.

[38] J. Bruckbauer, P. R. Edwards, S.-L. Sahonta, F. C. P. Massabuau, M. J. Kappers, C. J. Humphreys, R. A. Oliver, R. W. Martin, *J. Phys. D: Appl. Phys.* **2014,** *47*, 135107.

[39] A. Vaitkevičius, J. Mickevičius, D. Dobrovolskas, Ö. Tuna, C. Giesen, M. Heuken, G. Tamulaitis, *J. Appl. Phys.* **2014,** *115*, 213512.

[40] S. L. Rhode, W. Y. Fu, M. A. Moram, F. C. P. Massabuau, M. J. Kappers, C. McAleese, F. Oehler, C. J. Humphreys, R. O. Dusane, S. L. Sahonta, *J. Appl. Phys.* **2014,** *116*, 103513.

[41] T. J. O'Hanlon, F. C. P. Massabuau, A. Bao, M. J. Kappers, R. A. Oliver, *Ultramicroscopy* **2021,** *231*, 113255.

[42] F. C. P. Massabuau, L. Trinh-Xuan, D. Lodié, E. J. Thrush, D. Zhu, F. Oehler, T. Zhu, M. J. Kappers, C. J. Humphreys, R. A. Oliver, *J. Appl. Phys.* **2013,** *113*, 073505.

[43] G. Kusch, E. J. Comish, K. Loeto, S. Hammersley, M. J. Kappers, P. Dawson, R. A. Oliver, F. C. P. Massabuau, *Nanoscale* **2022,** *14*, 402.

[44] T. Zywietz, J. Neugebauer, M. Scheffler, *Appl. Phys. Lett.* **1998,** *73*, 487.

[45] Y.-H. Cho, S. K. Lee, H. S. Kwack, J. Y. Kim, K. S. Lim, H. M. Kim, T. W. Kang, S. N. Lee, M. S. Seon, O. H. Nam, Y. J. Park, *Appl. Phys. Lett.* **2003,** *83*, 2578.

[46] Z. Z. Chen, P. Liu, S. L. Qi, K. Xu, Z. X. Qin, Y. Z. Tong, T. J. Yu, X. D. Hu, G. Y. Zhang, *J. Cryst. Growth* **2007,** *298*, 731.

[47] Y. Feng, Z. Chen, C. Li, Y. Chen, J. Zhan, Y. Chen, J. Nie, F. Jiao, X. Kang, S. Li, Q. Wang, T. Yu, G. Zhang, B. Shen, *RSC Adv.* **2018,** *8*, 16370.

[48] Y.-C. Leem, S.-Y. Yim, *ACS Photonics* **2018,** *5*, 1129.

[49] Y.-S. Yoo, T.-M. Roh, J.-H. Na, S. J. Son, Y.-H. Cho, *Appl. Phys. Lett.* **2013,** *102*, 211107.

[50] C. C. Li, J. L. Zhan, Z. Z. Chen, F. Jiao, Y. F. Chen, Y. Y. Chen, J. X. Nie, X. N. Kang, S. F. Li, Q. Wang, G. Y. Zhang, B. Shen, *Opt Express* **2019,** *27*, A1146.

[51] S. Karpov, *Opt. Quantum Electron.* **2014,** *47*, 1293.

[52] L. Wang, L. Wang, C. J. Chen, K. C. Chen, Z. Hao, Y. Luo, C. Sun, M. C. Wu, J. Yu, Y. Han, B. Xiong, J. Wang, H. Li, *Laser Photonics Rev.* **2021,** *15*, 2000406.

[53] A. Hangleiter, F. Hitzel, C. Netzel, D. Fuhrmann, U. Rossow, G. Ade, P. Hinze, *Phys. Rev. Lett.* **2005,** *95*, 127402.

[54] F. A. Ponce, S. Srinivasan, A. Bell, L. Geng, R. Liu, M. Stevens, J. Cai, H. Omiya, H. Marui, S. Tanaka, *Phys. Status. Solidi. (b)* **2003,** *240*, 273.

[55] T. Sugahara, M. Hao, T. Wang, D. Nakagawa, Y. Naoi, K. Nishino, S. Sakai, *Jpn. J. Appl. Phys.* **1998,** *37*, L1195.

[56] J. Zhan, Z. Chen, C. Deng, F. Jiao, X. Xi, Y. Chen, J. Nie, Z. Pan, H. Zhang, B. Dong, X. Kang, Q. Wang, Y. Tong, G. Zhang, B. Shen, *Nanomaterials (Basel)* **2022,** *12*, 3880.

# Supporting Information

**Efficient InGaN-based Red Light-Emitting Diodes by Modulating Trench Defects**

*Zuojian Pan, Zhizhong Chen,\* Haodong Zhang, Han Yang, Chuhan Deng, Boyan Dong, Daqi Wang, Yuchen Li, Hai Lin, Weihua Chen, Fei Jiao, Xiangning Kang, Chuanyu Jia, Zhiwen Liang, Qi Wang,\* Guoyi Zhang, Bo Shen*

Z. Pan, Z. Chen, H. Zhang, H. Yang, C. Deng, B. Dong, D. Wang, Y. Li, H. Lin, W. Chen, F. Jiao, X. Kang, G. Zhang, B. Shen
State Key Laboratory for Artificial Microstructure and Mesoscopic Physics
School of Physics
Peking University
Beijing 100871, China
E-mail: zzchen@pku.edu.cn

Z. Chen, Z. Liang, Q. Wang, G. Zhang
Dongguan Institute of Optoelectronics
Peking University
Dongguan 523808, China
E-mail: wangq@pku-ioe.cn

Z. Chen, B. Shen
Yangtze Delta Institute of Optoelectronics
Peking University
Nantong 226000, China

F. Jiao
State Key Laboratory of Nuclear Physics and Technology
School of Physics
Peking University
Beijing 100871, China

C. Jia
School of International Academy of Microelectronics
Dongguan University of Technology
Dongguan 523000, China

**Figure S1. Effect of growth temperatures on the trench defect densities**

A series of experiments were conducted to demonstrate the effect of quantum well (QW) / quantum barrier (QB) temperatures on the formation of trench defects in green multi-quantum wells (MQWs), as shown in Figure S1. While varying the following QW/QB temperatures, the TMIn flux was adjusted to keep the peak wavelength at 510 nm, excluding the effect of In components on the trench defect densities. Results show that individually reducing either the QW or QB temperatures can result in higher trench defect density. Nonetheless, to achieve trench defect densities exceeding $10^9$ cm$^{-2}$, a simultaneous decrease in both QW and QB temperatures is necessary.

Previous studies show that the formation of trench defects is related to the In-rich clusters in QWs and the low Ga adatom mobility during the QB growth[29, 44]. When keeping the QW temperatures relatively high (700 °C), there are fewer In-rich clusters in the QWs. In this case, even if the QB temperature is lowered to same with the QW temperature, the trench defect density remains limited to $9.3 \times 10^7$ cm$^{-2}$. Similarly, when QB temperatures are kept relatively high (820 °C), the diffusion of Ga adatoms during the QB growth becomes stronger. Under this condition, by reducing the QW temperature to 630 °C, the trench defect density can only reach $2.7 \times 10^8$ cm$^{-2}$. Once the QW and QB temperatures are decreased in parallel, a significant increase in the trench defect density can be observed. Specifically, at QW/QB temperatures of 630 °C /700 °C, the trench defects density can reach $4.5 \times 10^9$ cm$^{-2}$. It means that high-density trench defects are easily formed when the In-rich clusters in QWs are combined with the low Ga-adatom mobility during QB growth.

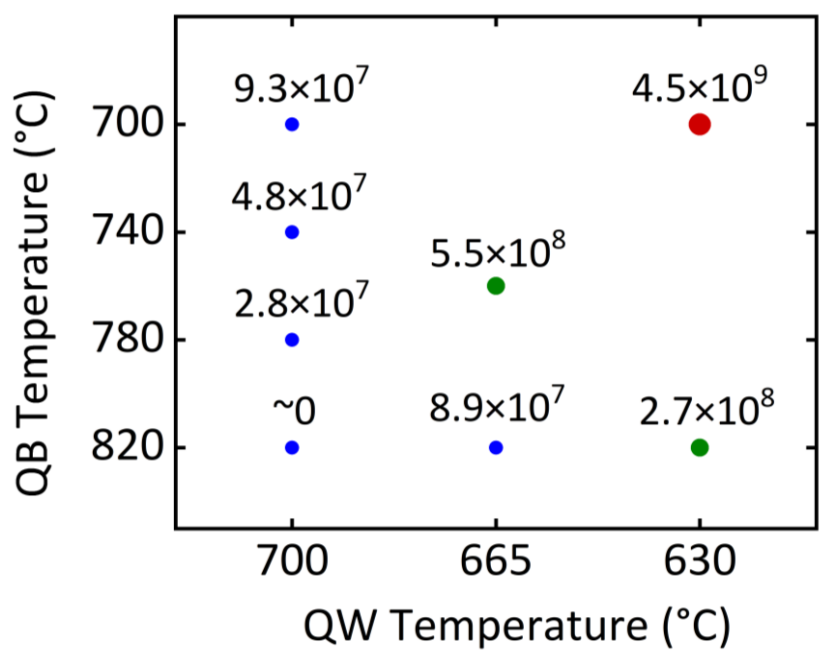

Figure S1. Variation of trench defect densities (cm$^{-2}$) with QW/QB growth temperatures of green MQWs.

**Figure S2. The surface morphology and emission properties of green MQWs**

The density of trench defects on the surface of green MQWs samples A1, B1, and C1 are 0, $(5.5\pm0.5)\times10^8$ cm$^{-2}$, and $(4.5\pm0.3)\times10^9$ cm$^{-2}$, respectively. Comparing the scanning electron microscopy (SEM) with cathodoluminescence (CL) images, the green MQWs within most trench defects exhibit dimmer luminescence. This decrease in CL intensity is mainly associated with the stacking mismatch boundaries (SMBs) in trench defects, which act as nonradiative recombination centers[33].

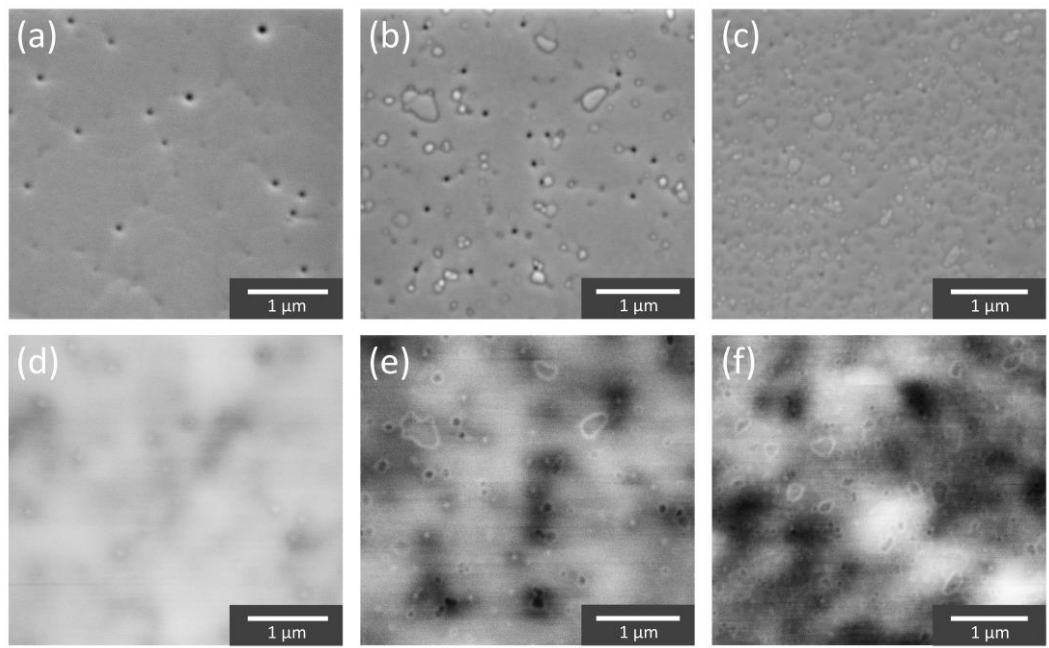

Figure S2. (a-c) SEM images of green MQWs samples A1, B1, and C1. (d-f) The corresponding monochromatic CL images at 500 nm.

**Figure S3. Direct observation of luminescence of the MQWs with various trench defect densities**

Since the 405 nm laser can excite both green and red MQWs, the image observed in the fluorescence microscope appears as a composite of luminescence images originating from the green and red MQWs. The strong green luminescence in Figure S3a mainly originates from the green MQWs of sample A3 due to their high green MQWs growth temperature. Meanwhile, the red MQWs also emit green light since the In component is relatively low. Subsequently, as the growth temperature of the green MQWs decreases, the green luminescence in samples B3 and C3 exhibits a significant decrease in intensity. Figure S3b-c illustrates that the red luminescence mainly originates from the dot-like emissions inside the trench defects. In Figure S3c, the luminescence predominantly originates from the red MQWs in sample C3, with considerably weaker green luminescence. These results coincide with CL and confocal photoluminescence (PL) findings.

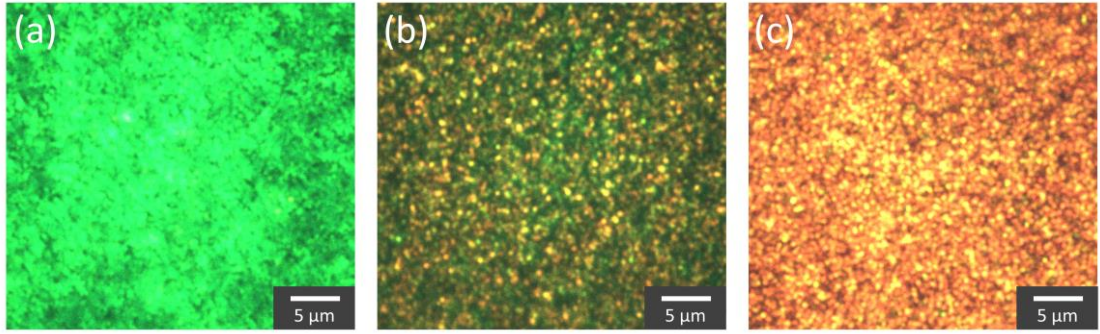

Figure S3. Fluorescence images of samples (a) A3, (b) B3, and (c) C3 taken under identical conditions with a 405nm laser.

**Figure S4. The surface morphology of red MQWs**

The trench defect densities in red MQWs of samples A3, B3, and C3 are $(5.7\pm0.5)\times10^8 cm^{-2}$, $(1.2\pm0.3)\times10^9 cm^{-2}$, and $(5.3\pm0.3)\times10^9 cm^{-2}$, respectively. The trench defect densities observed on the surface of the red MQWs is comparatively higher than that of the green MQWs. It means additional trench defect generation during the growth process of the red MQWs, mainly associated with low growth temperatures[23].

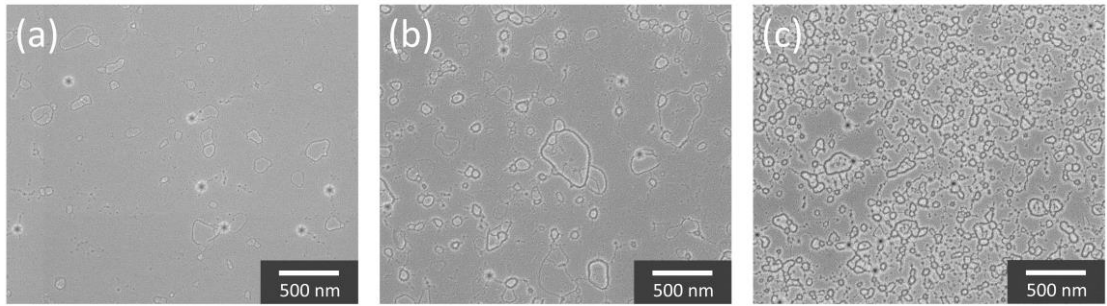

Figure S4. SEM images of the red LED samples (a) A3, (b) B3, (c) C3 without p-GaN layer.

**Figure S5. Confirmation of the higher In component in the red MQWs inside trench defects**

Figure S5a shows the cross-sectional high-angle annular dark field scanning transmission electron microscopy (STEM-HAADF) image of a type-A trench defect. STEM combined with energy-dispersive X-ray spectroscopy (STEM-EDX) was applied to investigate the In component distribution of MQWs inside and outside the type-A trench defects. Figure S5b-c demonstrates a gradual increase in the In component along the growth direction within the red MQWs inside the trench defects. In contrast, the In component within the red MQWs outside the trench defects remains relatively constant. The In components in the topmost red QW inside type-A trench defects is about $(32.0\pm3\%)$, which is about 7% higher than that in the surrounding region. The In component of blue QWs inside the trench defects was also observed to be 4% higher than the surrounding region in other studies[41]. The higher In component observed within the red MQWs inside the trench defects can be attributed to the increased In incorporation resulting from strain relaxation[14]. These findings align with the observations from CL and confocal PL, which reveal that the red MQWs inside the trench defects exhibit longer peak wavelengths compared to the surrounding region.

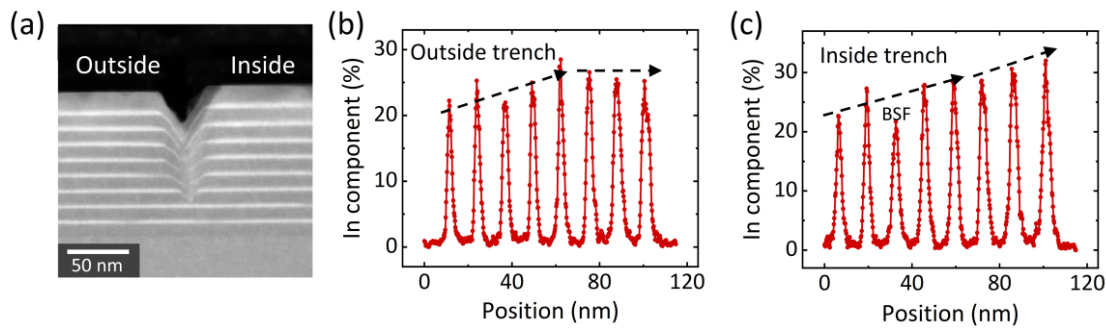

Figure S5. (a) STEM-HAADF image of a type-A trench defect. STEM-EDX profiles showing the variation in the atomic fraction of indium, taken from EDX maps across the QW stack (b) outside trench defect, (c) inside trench defect. The dashed line with arrows illustrates the observed trend of the In components.

**Figure S6-S9. Evidence of luminescence enhancement originating from the region inside trench defects intentionally introduced in the underlying green MQWs**

Introducing trench defects in the underlying green MQWs can significantly improve the luminescence efficiency of red MQWs, as confirmed by power-dependent PL and temperature-dependent PL in Figure 3. Further, the luminescence enhancement of the red MQWs mainly originates from the region inside the trench defects, as demonstrated by CL and confocal PL results in Figure 4-6. By comparing the SEM-CL images in Figure 4, it is evident that only a part of the trench defects exhibits luminescence enhancement and wavelength redshift.

The correlation between the structural features of trench defects and the emission properties of their internal red MQWs has been investigated, as shown in Figure S6-S9. Trench defects are classified by the width of the trench, the area inside the trench, and the prominence above surroundings[42]. As shown in Figure S6a, trench defects are classified and characterized as follows: type-A (wide trench, level center, large area); type-B (narrow trench, level center, large area); type-C (wide trench, prominent center, small area); type-D (narrow trench, level center, small area); type-E (concave center). Figure S6b-c displays the 3D atomic force microscopy (AFM) images of the type-A and type-C trench defects.

For each type of trench defect, CL line scannings are applied to investigate the luminescence properties of their internal red MQWs. Figure S7 shows the AFM line profiles and the corresponding CL line scanning results. It can be observed that luminescence enhancement and wavelength redshift only occur inside trench defects with wide and deep trenches, including type-A and type-C ones. The red MQWs inside the type-A trench defects exhibit an integrated intensity enhancement of about 3.0 times and a wavelength redshift of more than 20 nm compared to the surrounding region. Meanwhile, the red MQWs inside the type-C trench defects exhibit an integrated intensity enhancement of about 1.5 times and a wavelength redshift of 10 nm compared to the surrounding region. No obvious intensity enhancement or wavelength redshift is observed for red MQWs inside v-pits and other types of trench defects.

Figure S8 displays the dark-field transmission electron microscopy (TEM) images corresponding to each type of trench defect. Cross-sectional TEM specimens regarding trench defects were prepared on focused ion beam (FIB)-SEM system. A marking method of electron beam carbon coating was employed to position specific trench defects. The dark-field TEM results show that for type-A and type-C trench defects with deep trenches, their basal plane stacking fault (BSFs) originate from the underlying green MQWs. In other words, type-A and type-C trench defects are introduced intentionally in the underlying green MQWs. Notably, the red MQWs inside the type-A and type-C trench defects are spatially isolated from the underlying BSFs and SMBs, which could contribute to reducing defect-related recombination. For other types of trench defects, the BSFs and SMBs are located near the red MQWs, adversely affecting luminescence.

Regarding the type-A and type-C trench defects with deep grooves, the inner region of the type-A trench defect is nearly flat compared to the surrounding region, while the type-C trench defect has a prominence of about 15 nm. The CL line scanning results show that the luminescence enhancement inside the type-C trench defects is relatively lower than that of the type-A ones. Figure S9 shows the STEM-HAADF images of the type-A and type-C trench defects. The results show that the QW widths inside the type-A trench defects are basically the same as those in the surrounding region. In contrast, the QW width inside the type-C trench defects increases monotonously from 2.47 nm to 5.62 nm along the growth direction. The wider QW width will be detrimental to the luminescence of the red MQWs due to the larger polarization electric field and weaker quantum confinement of carriers [9].

In summary, through a detailed investigation of the structural characteristics of trench defects and the emission properties of their internal red MQWs, we have confirmed that the deep trench defects, which can enhance luminescence, originate from intentional introduction within the underlying green MQWs. For trench defects with deep grooves, more prominence generally corresponds to a smaller enhancement.

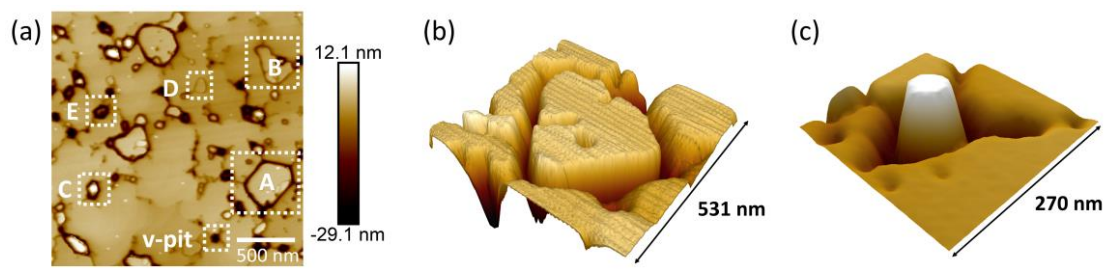

Figure S6. (a) AFM image of the full-structure sample B3 without p-GaN layer. The dashed rectangle shows the typical microstructures that appear on the surface of red MQWs. 3D AFM images of two types of trench defects exhibiting luminescence enhancement and wavelength redshift (b) a type-A trench defect. (c) a type-C trench defect.

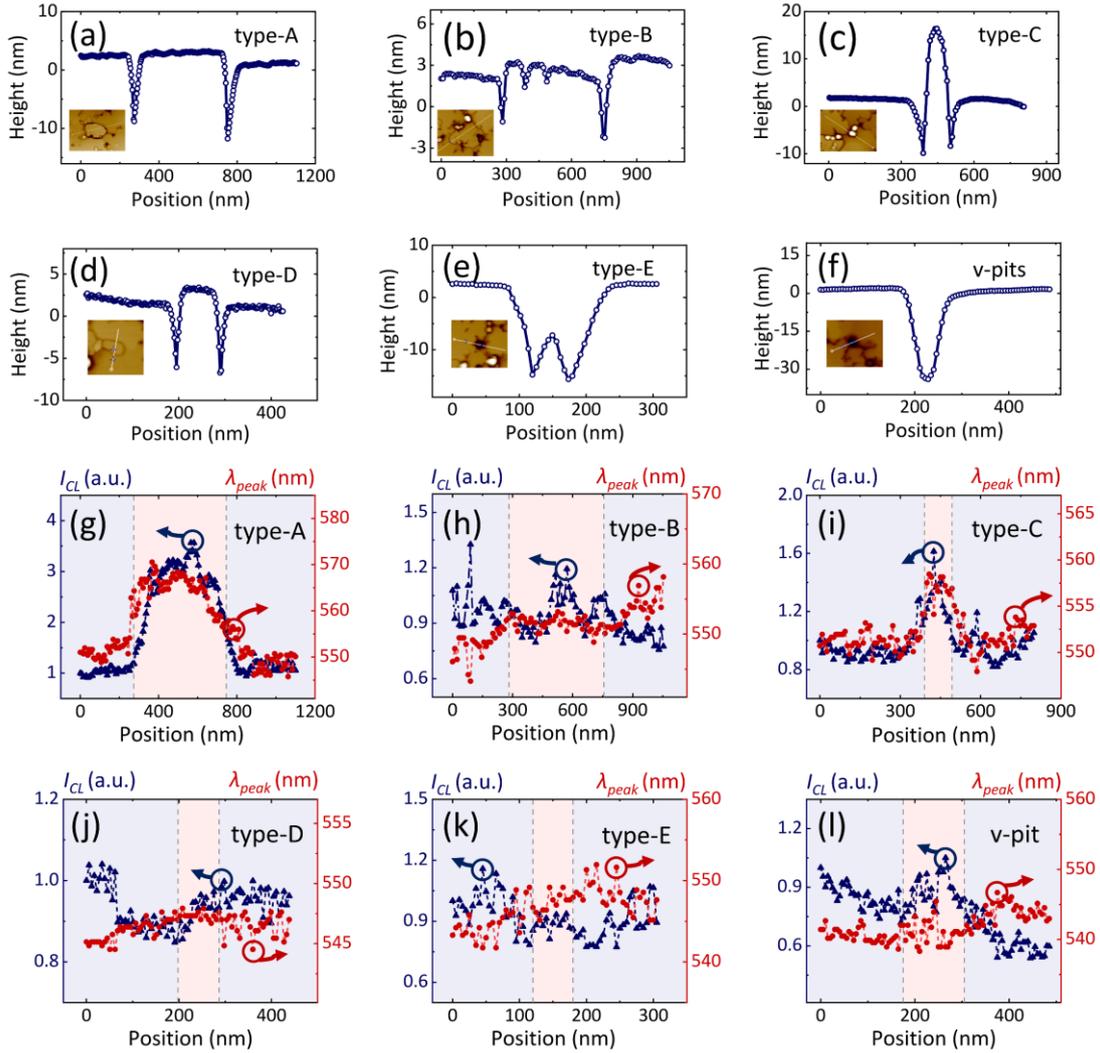

Figure S7. AFM line profiles of (a) type-A, (b) type-B, (c) type-C, (d) type-D, (e) type-E trench defects, and (f) v-pit on the surface of red MQWs in sample B3. The insets show the AFM images and the white lines mark the positions of the AFM line scans. The corresponding CL integrated intensity and peak wavelength distributions obtained from CL line scans across (g) type-A, (h) type-B, (i) type-C, (j) type-D, (k) type-E trench defects, and (l) v-pit. The red dotted lines represent the peak wavelength distributions, and the blue dotted lines represent the CL integrated intensity distributions. The reddish areas represent the regions inside trench defects or v-pits, and the bluish areas represent the regions surrounding these surface structures. Note that the AFM line scans in (a-f) and the CL line scans in (g-l) are in the same position. The AFM and SEM/CL images of the same area are positioned by special markers.

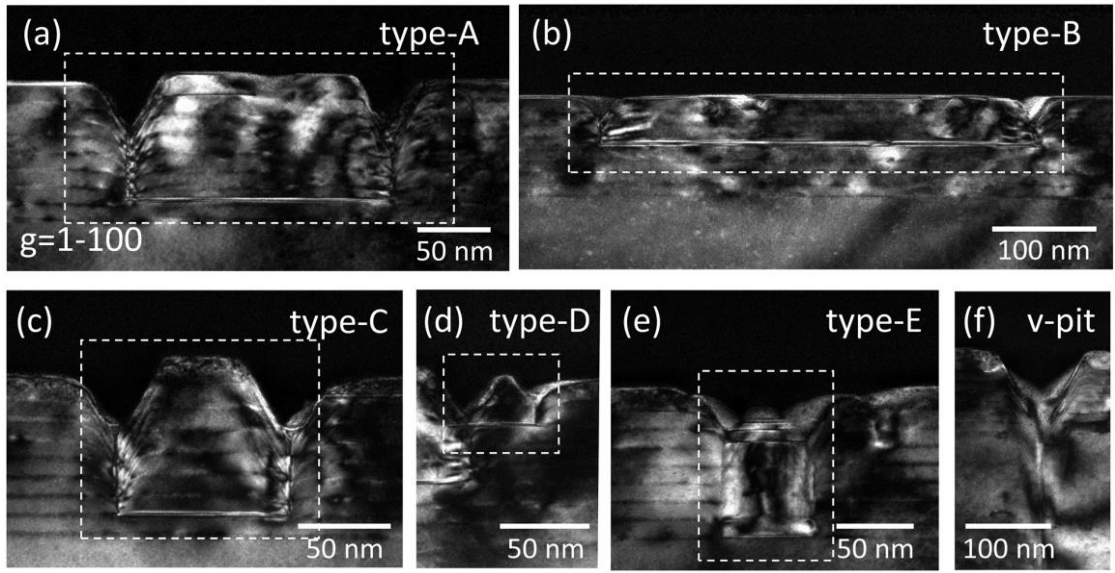

Figure S8. Cross section of (a) type-A, (b) type-B, (c) type-C, (d) type-D, (e) type-E trench defects, and (f) v-pit in dark-field TEM (zone axis 11-20) imaging mode with g (1-100). The dashed rectangles mark the location of the trench defects.

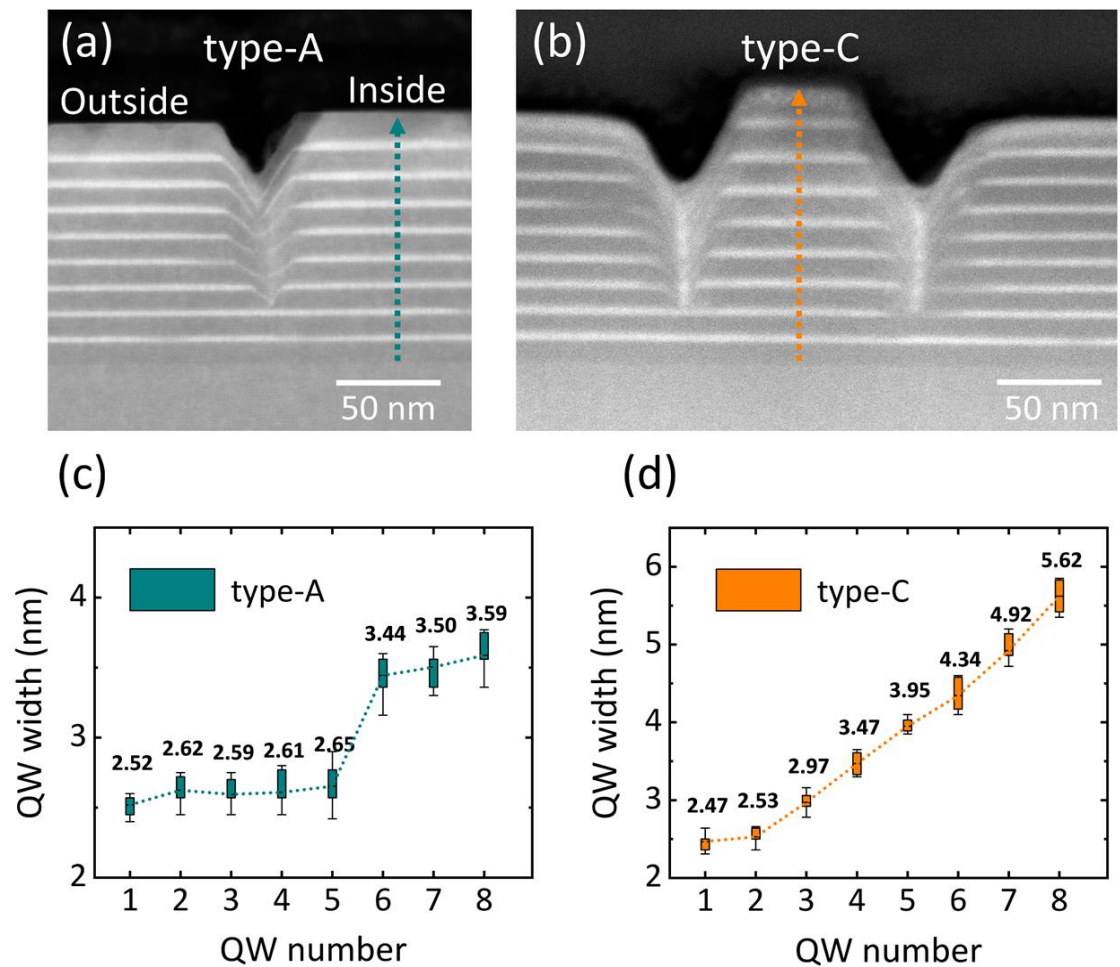

Figure S9. STEM-HAADF images of (a) type-A trench defect. (a) type-C trench defect. The dashed lines with arrows indicate the location and direction of measuring the QW

widths. Variation of QW widths inside (c) type-A trench defect, (d) type-C trench defect, measured from STEM-HAADF images, with QW numbers from 1 to 8 ordered along the growth direction.

**Figure S10. Effect of shallow trench defects on red MQWs with low In component**

The low growth temperatures of the red MQWs result in the new formation of shallow trench defects with density of $(5.7\pm0.5)\times10^8\,cm^{-2}$. The BSFs and SMBs of these trench defects are embedded in the red MQWs, which is detrimental to the luminescence. The red MQWs inside trench defects exhibit weaker luminescence due to embedded defects such as BSFs and SMBs, as shown in Figure S10. The luminescence enhancement at the edge of trench defects can be attributed to the improved light extraction efficiency. These observation coincides with the low internal quantum efficiency of the red QWs in sample A2.

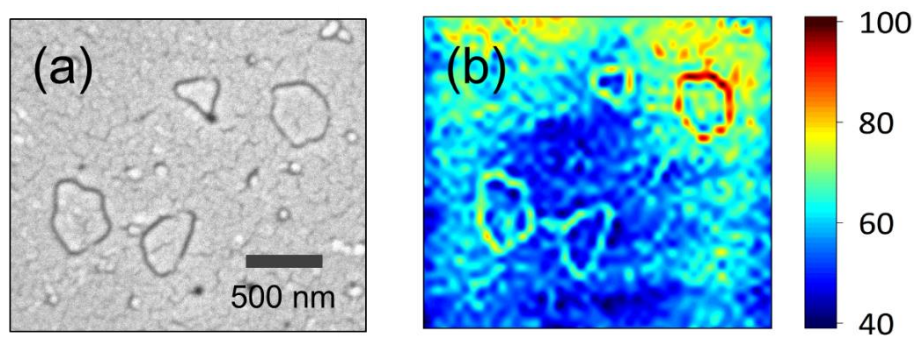

Figure S10. (a) SEM image of the MQWs-interrupted sample A2, (b) The corresponding CL integral intensity maps of red MQWs after split-peak fitting.

**Figure S11. Measurement of stress states in MQWs outside trench defects**

X-ray diffraction reciprocal space mappings (XRD RSMs) were applied to evaluate the stress state of MQWs. Figure S11 shows the XRD RSM images along the (105) reflections of samples A2, B2, and C2. Notably, the satellite peaks measured by XRD RSM all originate from the MQWs outside trench defects, not reflecting the stress state of the MQWs inside trench defects. To obtain satellite peaks in XRD RSM, it is necessary that the MQWs have flat interfaces, fixed periods and In components. Due to the different prominences (-10nm ~ +20nm) and varying peak wavelengths, those MQWs inside trench defects have different periods and In components. Therefore, no signals from the MQWs inside trench defects can be seen in the XRD RSM. In the Figure S11 below, all the satellite peaks originate from the underlying green MQWs and the red MQWs outside trench defects. Moreover, from the vertical alignment of satellite peaks, these MQWs are fully strained with the underlying GaN layer.

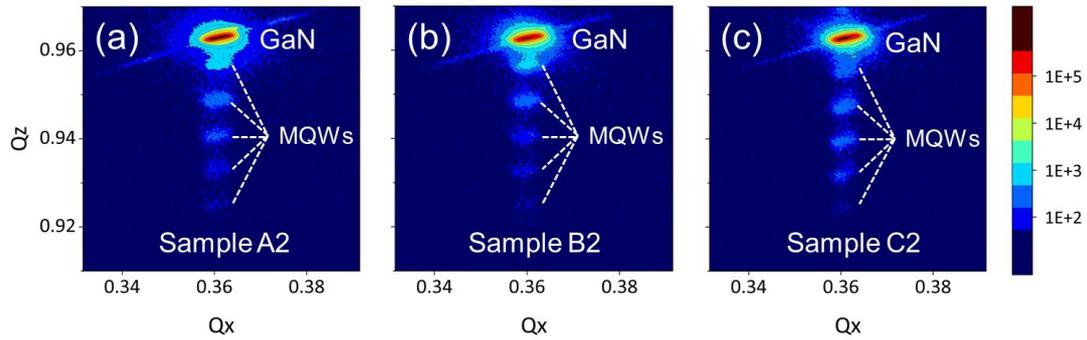

Figure S11. High-resolution XRD RSMs along the (105) reflections of the full-structure red LED samples (a) A2, (b) B2, and (c) C2.

**Figure S12. Simulation of the reason for luminescence enhancement of red MQWs inside trench defects**

To clarify the effects of polarization, Shockley-Read-Hall (SRH) recombination and Auger recombination on the red MQWs inside and outside the trench defects, simulations were performed using Crosslight software[52]. Specifically, an optical pumping model of red MQWs was employed to simulate confocal PL experiments with varying laser power densities. Considering that the 532 nm laser cannot excite the green MQWs, the full-structure red InGaN LED model was simplified, as depicted in Figure S12a. The critical differences between the red MQWs inside and outside of the trench defects include the In component, the internal polarized electric field, the SRH recombination, and the Auger recombination.

Two structures, LED_A and LED_B, were employed to simulate the MQWs inside and outside the trench defects, respectively. For the LED_A structure, the In component of red MQWs was set at 33.3%, the polarization coefficient was set as 0.47, the SRH lifetime was set as $5.0 \times 10^{-8}$ s, and the Auger lifetime was set as $1.2 \times 10^{-6}$ s. For the LED_B structure, the In component of red MQWs was set at 25.5%, the polarization coefficient was set as 0.85, the SRH lifetime was set as $2.2 \times 10^{-9}$ s, and the Auger lifetime was set as $7.5 \times 10^{-7}$ s. The polarization coefficients range between 0 and 1, where the closer to 1, the stronger the polarized electric field is in the MQWs. The remaining parameters of the LED_A and LED_B structures remain consistent. These parameters are set based on the fact that the MQWs inside the trench show a higher In component, a weaker polarized electric field, less SRH recombination, and more Auger recombination compared to those outside the trench. In order to figure out the influences of polarization, SRH recombination, and Auger recombination on luminescence, LED_A was adjusted to the same Auger lifetime as LED_B and defined as LED_A', and LED_B was adjusted to the same SRH lifetime as LED_A and defined as LED_B'. By comparing the integral intensities of LED_A, LED_A', LED_B, and LED_B' at different laser power densities, the proportion of polarization effect, SRH recombination, and Auger recombination influencing luminescence can be qualitatively judged.

The simulation results show that as the laser power density is varied by three orders of magnitude, the peak wavelength of LED_A is shifted from 599 nm to 570 nm, and the peak wavelength of LED_B is shifted from 608 nm to 541 nm, which is consistent with the experimental results. Here, due to the spatial resolution limitation of confocal PL, the peak wavelengths of red MQWs at high injection are based on CL measurement. Figure S12b illustrates the energy band diagrams of the red MQWs at high injection inside and outside trench defects, indicating a weaker polarization electric field in the MQWs inside the trench defects. Figure S12c shows the integral intensity variation of the MQWs inside and outside trench defects corresponding to variable laser power. At high injection, the polarization effect in red MQWs outside trench defects is the primary factor contributing to weaker luminescence than inside trench defects. Nevertheless, SRH recombination serves as the predominant factor for the drastic decrease in the radiative recombination efficiency of the red MQWs outside the trench defects at low injection. At higher injection levels, the gap in luminescence between the inside and outside of trench defects can be predicted to decrease further.

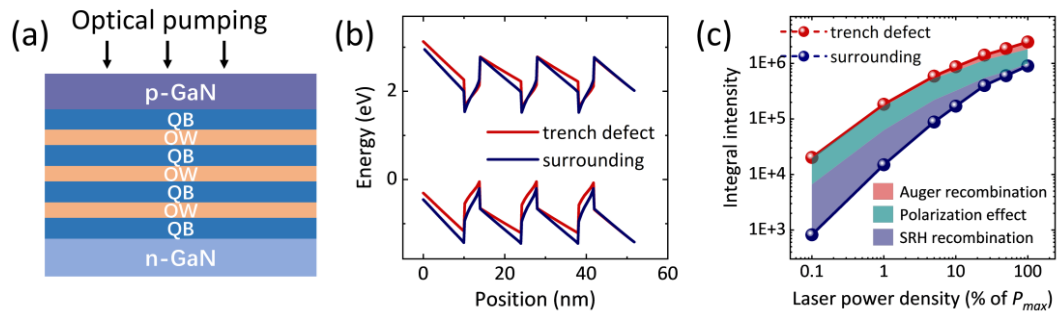

Figure S12. (a) Schematic of the structural model in the simulation. (b) Energy band diagrams of the MQWs at the laser power density of $3.1 \times 10^6$ W/cm$^2$ inside and outside trench defects obtained from simulations. (c) Dependencies of PL integral intensity of the MQWs inside and outside trench defects on the laser power obtained by simulation. The $P_{max}$ corresponds to $3.1 \times 10^6$ W/cm$^2$.

**References**
[1] T. Zywietz, J. Neugebauer, M. Scheffler, *Appl. Phys. Lett.* **1998,** *73*, 487.
[2] Z. Shi, A. Tian, X. Sun, X. Li, H. Zang, X. Su, H. Lin, P. Xu, H. Yang, J. Liu, D. Li, *J. Appl. Phys.* **2023,** *133*, 123103.
[3] F. C. P. Massabuau, S. L. Sahonta, L. Trinh-Xuan, S. Rhode, T. J. Puchtler, M. J. Kappers, C. J. Humphreys, R. A. Oliver, *Appl. Phys. Lett.* **2012,** *101*, 212107.
[4] F. C. P. Massabuau, A. Le Fol, S. K. Pamenter, F. Oehler, M. J. Kappers, C. J. Humphreys, R. A. Oliver, *Phys. Status. Solidi. (a)* **2014,** *211*, 740.
[5] T. J. O'Hanlon, F. C. P. Massabuau, A. Bao, M. J. Kappers, R. A. Oliver, *Ultramicroscopy* **2021,** *231*, 113255.
[6] G. B. Stringfellow, *J. Cryst. Growth* **2010,** *312*, 735.
[7] F. C. P. Massabuau, L. Trinh-Xuan, D. Lodié, E. J. Thrush, D. Zhu, F. Oehler, T. Zhu, M. J. Kappers, C. J. Humphreys, R. A. Oliver, *J. Appl. Phys.* **2013,** *113*, 073505.
[8] V. Fiorentini, F. Bernardini, F. Della Sala, A. Di Carlo, P. Lugli, *Phys. Rev. B* **1999,** *60*, 8849.
[9] APSYS, (2023 version) by Crosslight Software, Inc., Burnaby, Canada, http://www.crosslight.com.